\documentclass[a4paper,12pt]{article}
\usepackage{soul}
\usepackage[usenames,dvipsnames]{color}
\usepackage{jheppub}
\usepackage{amsmath, amssymb, slashed, epsf, color, graphicx, latexsym}
\usepackage{epsfig}
\usepackage{graphics}

\usepackage{mathtools}% http://ctan.org/pkg/mathtools
\newcommand{\nn}{\nonumber}

\begin{document}
	
%	\preprint{get one}
	\title{Action complexity of charged black holes with higher derivative interactions}

	\author[a]{Taniya Mandal, }
	\author[b] {Arpita Mitra, }
	\author[c] {and Gurmeet Singh Punia}
	%\author[d] {}
	
	\affiliation[a]{National Institute for Theoretical Physics, School of Physics and Mandelstam Institute for Theoretical Physics, University of the Witwatersrand, Wits, 2050, South Africa}
	\affiliation[b]{Physics Division, National Center for Theoretical Sciences, Taipei 10617, Taiwan}
	\affiliation[c]{Indian Institute of Science Education and Research Bhopal, Bhopal Bypass, Bhopal 462066, India}

	\emailAdd{taniya.mandal@wits.ac.za}
	\emailAdd{arpitamitra89@gmail.com}
	\emailAdd{gurmeet17@iiserb.ac.in}
	
	\abstract{Quantum complexity of CFT states can be computed holographically from the dual gravitational solutions. In this paper, we have studied the late time growth of holographic complexity of a charged black hole in five dimensional Anti-de Sitter spacetime in the presence of quartic derivative interaction terms using the Complexity = Action conjecture. These interaction terms in the gravitational action can lead to the violation of Llyod's bound. The dual CFT is known to admit a hydrodynamic description where the KSS bound is also violated due to the presence of higher derivative terms in the bulk action. The origin of terms which violate both the bounds are the same for the gravitational action of consideration. We have also discussed the late time complexity growth rate of the Jackiw-Teitelboim (JT) model with higher derivative corrections. 
	}

%opening

\maketitle
%\tableofcontents
\section{Introduction}
The AdS/CFT conjecture provides a non-perturbative formulation of quantum gravity in asymptotically Anti-de Sitter (AdS) spacetime in terms of a field theory living on the boundary \cite{Maldacena:1997re}. One interesting direction to explore is  how spacetime emerges from field theory degrees of freedom within the AdS/CFT correspondence \cite{VanRaamsdonk:2010pw}. Aspects of quantum information on the boundary CFT are manifest through geometric quantities in the bulk.  
One such example of geometrization is the holographic Ryu-Takayanagi prescription for static bulk duals. The entanglement entropy of a subregion in the CFT is dual to the area of a co-dimension two, minimal surface anchored on the boundary of the subregion \cite{Ryu:2006ef}. For a $d$-dimensional CFT the entanglement entropy is,
\begin{equation}
    S_{EE}=\frac{\text{min}({\cal{A}})}{4G_N^{(d+1)}},
\end{equation}
where ${\cal{A}}$ is the area functional and $G_N^{(d+1)}$ is the Newton constant in the bulk.  For static backgrounds we have a unique minimum while if we consider time dependent boundary states the area functional may have multiple saddle points. Therefore for time dependent cases one should follow the covariant generalization of the holographic entanglement entropy, usually known as the Hubeny-Rangamani-Takayanagi (HRT) construction \cite{Hubeny:2007xt} where one needs to consider the minimum amongst the extremal surfaces,
\begin{equation}
    S_{EE}=\frac{\text{min}(\text{extremal}~{\cal{A}})}{4G_N^{(d+1)}}.
\end{equation}
Another interesting observation which motivated a connection between quantum information theory with geometry through the holographic duality is the relation of entropy with the number of degrees of freedom in the dual quantum description of a black hole,
\begin{align}
    n_{\text{microstates}}=e^{S_{\text{bh}}}
\end{align}
where  $ n_{\text{microstates}}$ is the dimension of Hilbert space. 

However, entanglement entropy is not the appropriate quantity to describe the evolution of Einstein-Rosen bridge in the interior of a black hole \cite{Susskind:2014moa}. The wormhole which connects the two sides of an eternal AdS-Schwarzschild black hole grows with time still after it reaches thermal equilibrium i.e. the growth of ER bridge continues for a much longer time compared to the thermalization time. It can be realized by measuring the volume of a spacelike slice stretching through the wormhole \cite{Susskind:2014rva, Stanford:2014jda}. %New quantity was required to express the CFT dual of this growth. 
Recent progress in the interpretation of spacetime geometry in terms of quantum information theoretic quantities is improving our understanding of a complete holographic description of the black hole interior.  Since in quantum information theory the complexity of a state can increase under time evolution, in recent times a conjecture was proposed which relates the increase of complexity with the late time growth of the black hole interior. Therefore one can state that quantum computational complexity of the boundary field theory state is encoded geometrically in the dual gravitational spacetime. While the entropy is related to the loss of information, complexity is related to the difficulty to process that information.

There are two conjectures which were proposed to compute the holographic complexity. One of these are the Complexity=Volume (CV) conjecture, which states that the complexity is proportional to the volume of a maximal codimension-1 slice anchored at the boundary  \cite{Susskind:2014rva, Stanford:2014jda}
\begin{equation}
C_V \sim \frac{\text{Max(V)}}{Gl},
\end{equation}
where $l$ is a length scale associated with the bulk we are considering. The proportionality factor depends on the specifics of the black hole. The other conjecture is known as the Complexity=Action (CA) conjecture, where the complexity of boundary states is proportional to the gravitational action evaluated in the Wheeler-De Witt (WDW) patch, i.e. the region bounded by the null surfaces anchored at left and right boundary at a chosen time \cite{Brown:2015bva, Brown:2015lvg, Goto:2018iay},
\begin{equation}
C_A =\frac{I_{\text{WDW}}}{\pi\hbar},
\end{equation}
Another recent proposal has been suggested to define the holographic dual of complexity, where complexity is proportional to the spacetime volume of the WDW patch. This proposal is known as C$_\text{V}^{2.0}$ conjecture in the literature and is succinctly expressed through the formula \cite{An:2018dbz,Couch:2016exn}. 
\begin{equation}
    C_{\text{V}}^{2.0}=\frac{V_{\text{WDW}}}{Gl^2}.
\end{equation}
Recently in \cite{Belin:2021bga}, as possible candidates for a gravitational dual of complexity an infinite family of gravitational observables on codimension-one slices of the  geometry is introduced.   

Holographic complexity has been recently studied in various asymptotically AdS backgrounds \cite{Alishahiha:2015rta, Brown:2016wib, Chapman:2016hwi, Reynolds:2016rvl, Cai:2016xho, Lehner:2016vdi, Carmi:2017jqz, Bolognesi:2018ion} and also in their deformations \cite{Ghodrati:2017roz, Auzzi:2018zdu, Auzzi:2018pbc, Auzzi:2019fnp, Auzzi:2021nrj}. It was also computed in the presence of defects and boundaries \cite{Chapman:2018bqj, Braccia:2019xxi, Sato:2019kik, Baiguera:2021cba, Auzzi:2021ozb, Sato:2021ftf}. The recent advancements of holographic complexity in the asymptotic de Sitter space has been discussed in \cite{Susskind:2021esx, Chapman:2021eyy, Jorstad:2022mls}. One of the key issues regarding these two conjectures is how to differentiate between CV and CA in the bulk dual. %One common feature between these two conjectures is that their late time behaviour agree.
The late time behaviour of the complexities arising from either conjectures are in agreement. In considering the late time limit of complexity, one can notice that if the computation is done using the CV conjecture, the complexity rate increases until %a time scale exponential in the number of degrees of freedom in the system and then 
it saturates to an upper bound  which satisfies the Llyod's bound \cite{Lloyd_2000}. Unlike the former case, the complexity evaluated using the CA conjecture increases and then overshoots this bound at intermediate times, until at late times it saturates the bound from above. While the CV conjecture successfully explains that the maximal volume grows like $TS$ ($T$=Temperature of the black hole, $S$=Entropy of the black hole), which is expected also for quantum computational complexity, the involvement of the length scale lacks the universal description. There is no first order principle to explain why the maximum volume slice is preferred and does not foliate the entire geometry behind the horizon at an instant of time. These deficiencies were cured in the CA conjecture where no length scale is involved and the gravitational action is computed in a region which contains the entire family of so-called ``nice slices" residing within the WDW patch.%null rays projected from both the boundaries at a chosen time.

In this paper we are studying the effect of generic, four-derivative  corrections to the growth of holographic complexity of a charged black hole in asymptotically AdS space in five spacetime dimensions, using the CA conjecture. Higher derivative corrections in general arise in the low energy effective theory of superstring theory \cite{Zwiebach:1985uq,Gross:1986iv,Gross:1986mw,Myers:1987yn}. A certain combination of higher curvature corrections to the Einstein gravity leads to the so-called Lovelock gravity, which is the most general metric theory of gravity in arbitrary number of spacetime dimensions. Even though the solution gets modified, the equations of motion still retain a two-derivative structure. In effect, the theory remains ghost free. The effect of higher curvature terms are studied extensively in several contexts. For a strongly coupled gauge theory dual to a gravity which include higher derivative corrections in asymptotically AdS space, the universal ratio (which exist for two derivative gravity dual)  of shear viscosity to entropy density gets modified \cite{Buchel:2008vz}. Thus in the presence of certain higher derivative terms the Kovtun-Son-Starinets (KSS) bound \cite{Kovtun:2004de} gets violated. 

Holographic complexity of black hole solutions with some specific higher derivative corrections either involving Riemann term or the electromagnetic field strength have recently been studied. In particular the complexity growth rate was computed for charged black holes in Lovelock gravity \cite{Cano:2018aqi, Cano:2018ckq}, dyonic black hole with quartic electromagnetic field strength corrections \cite{Razaghian:2020bfk}, Gauss Bonnet gravity \cite{ An:2018dbz}, general quadratic curvature gravity \cite{Ghodsi:2020qqb}. In this paper, we consider higher derivative quartic corrections not only in Riemann curvatures but also in electromagnetic field strength in five spacetime dimensions. Thus not only the black hole solution, the chemical potential also gets modified as well. To avoid the presence of unphysical ghost particles, we will treat the higher derivative terms perturbatively. The black hole we are considering is charged and the holographic complexity for charged black hole were considered in multiple works \cite{Goto:2018iay, Carmi:2017jqz}. Specifically the near extremal case was studied in \cite{Alishahiha:2019cib}. 

Holographic complexity satisfies an upper bound which is known as Lloyd’s bound \cite{Lloyd_2000}. It is well established that for neutral black holes, the Lloyd’s bound is equal to twice of the black hole mass. We also know that the Lloyd's bound continues to be valid in the presence of the higher order curvature corrections \cite{Cano:2018aqi, Cano:2018ckq} or momentum relaxation terms \cite{Babaei-Aghbolagh:2021ast}. For charged black holes, the Llyod's bound involves the chemical potential. However in the presence of non-linear terms involving dilatonic or gauge fields, one obtains additional correction terms, which may result in the violation of the conventional bound. The case of Riemann squared term and the interaction between the Riemann tensor and electromagnetic field strength tensor have not been considered yet. The main purpose of this paper is to understand whether the presence of these terms violate the bound. Another motivation is to gain insight regarding any possible relation between the KSS bound and the Llyod bound. In \cite{Myers:2009ij}, charged planar AdS black hole solutions in the presence of these specific corrections and the hydrodynamic properties of the plasma in the dual CFT were considered. It was demonstrated that the universal ratio relating the shear viscosity and shear diffusion of perturbations violate the KSS bound in a certain limit. Therefore we can comment about the nature of violation of both the bounds.

The two dimensional Jackiw-Teitelboim (JT) gravity is the simplest model of quantum gravity. JT gravity is dual to the $0+1$ dimensional Sachdev-Ye-Kitaev(SYK) model in the low energy limit. JT gravity also describes the near extremal behaviour of the higher dimensional black holes upon dimensional reduction. Following the CA conjecture, complexity growth of the two dimensional JT gravity has been discussed widely \cite{Alishahiha:2018swh,Brown:2018bms}. It has been shown that the late time complexity growth rate for the two dimensional dilatonic model is non vanishing only when we consider them upon dimensional reduction of higher dimensional action with the Maxwell boundary term. Complexity of JT gravity following CV conjecture and a comparison of that with the Krylov complexity of the dual SYK model has been discussed in \cite{Jian:2020qpp}. A higher derivative corrected JT like model has been discussed in \cite{Banerjee:2021vjy} which encapsulates the near extremal behaviour of a four dimensional black holes with arbitrary quartic corrections in four dimensions. In this paper, we also briefly discuss the late time complexity growth rate of this higher derivative corrected JT like model in two dimensions. 

The plan of the paper is as follows: In section \ref{sec1}, we discuss the gravitational action of our interest and the black hole solution of it. In section \ref{sec2} we compute the holographic complexity of the above mentioned gravitational solution following the CA conjecture. Complexity growth rate of the higher derivative corrected JT gravity are discussed in section \ref{sec3}. Finally we conclude in section \ref{sec4}.

\section{Charged black holes in most generic four derivative theory of gravity}\label{sec1}
The most generic four derivative theory of gravity coupled to $U(1)$ gauge field and in the presence of a negative cosmological constant in 4+1 spacetime dimensions is described by the following bulk action \cite{Myers:2009ij}, \cite{Cremonini:2019wdk},
\begin{eqnarray}\label{action}
\begin{aligned}
I&= \int d^5 x\sqrt{-g}{\cal L}\\&=\frac{1}{16 \pi G}\int d^5 x\sqrt{-g}\Bigg( R-\frac{1}{4} F^2 +\frac{12}{L^2}\\
&+\lambda \Big( a_1 R_{\mu\nu\alpha\beta}R^{\mu\nu\alpha\beta}+ a_2 R_{\mu\nu\alpha\beta}F^{\mu\nu}F^{\alpha\beta} + a_3 (F^2)^2+ a_4 F^4\Big)\Bigg),
\end{aligned}
\end{eqnarray}
where $F^2=F_{\mu\nu}F^{\mu\nu}$, $F^4=F^\mu_\nu F^\nu_\alpha F^\alpha_\beta F^\beta_\mu$. The action consists of the Einstein-Hilbert term, Maxwell term, four-derivative interactions  involving Riemann tensor and electromagnetic field strength. The coefficients of higher derivative terms $a_1, a_2, a_3$ and $a_4$ have dimensions of $[L^2]$. Whereas we set the parameter $\lambda$ which controls the strength of the higher derivative corrections to be dimensionless. We will consider the higher derivative interaction terms as a perturbative correction, thus $\lambda \ll1$.

We require a following boundary action to have a well defined variation of the bulk action (\ref{action}) with respect to the metric on spacelike or timelike boundary surfaces,
\begin{align}\label{bdryaction}
I_b&= \frac{1}{8 \pi G}\int d^4 x\sqrt{-h}K\notag\\&+\frac{\lambda}{8 \pi G}\int d^4 x\sqrt{-h}\Bigg[a_1 \Bigg(-\frac{2}{3}K^3+2KK_{ab}K^{ab}-\frac{4}{3}K_{ab}K^{bc}K_c^a-4\left(\mathcal{R}_{ab}-\frac{1}{2}\mathcal{R}h_{ab}\right)K^{ab}\Bigg)\notag\\&+ 8\frac{a_1}{L^2}K - \frac{5 a_1}{6}K F^2 + 2 a_1 (K n_\mu F^{\mu\lambda}n_\nu F^\nu_\lambda+K_{ab}F^{a\lambda}F^b_\lambda)+2a_2 n_\mu F^{\mu a}n_\nu F^{\nu b} K_{ab} \Bigg],
\end{align}
where $h_{ab}$ is the induced metric, $n_{\mu}$ is the unit normal, $K_{ab}$ is the extrinsic curvature with $K$ as it's trace and $\mathcal{R}_{ab}$ is the boundary Ricci tensor. Latin indices $a,b,\cdots$ run from 1 to 4 whereas Greek indices $\mu,\nu,\cdots$ run from 1 to 5. 

Similarly to have a well behaved variation with respect to the gauge field, we add the following boundary term to the gravitational action,
\begin{align}\label{gaugevariation}
    I_{\mu Q}=&\frac{\gamma}{16\pi G}\int d\Sigma_{\mu} F^{\mu\nu}A_{\nu}-\frac{\gamma}{4\pi G}\lambda a_2 \int d\Sigma_{\mu} R^{\mu\nu\alpha\beta} F_{\alpha\beta}A_{\nu}\notag\\&-\frac{\gamma}{2\pi G} \lambda a_3 \int d\Sigma_{\mu}F^2 F^{\mu\nu}A_{\nu}-\frac{\gamma}{2\pi G} \lambda a_4\int d\Sigma_{\mu} F^{\mu\gamma} F^{\mu\delta} F_{\gamma\delta}A_{\nu},
\end{align}
which does not affect the equations of motion but it modifies the boundary condition imposed on the gauge field. $\gamma$ is an arbitrary dimensionless coefficient.

The equation of motion attained by varying \eqref{action} with respect to the metric is,
\begin{align}\label{EH}
    R_{\mu\nu} &-\frac{1}{2} R g_{\mu \nu}=\frac{1}{2} F_{\mu \rho} F_{\nu}^{\rho}-\frac{1}{8} F^{2} g_{\mu\nu}+\frac{6}{L^{2}} g_{\mu\nu} \notag\\
    &+\lambda a_{1}\left(\frac{1}{2} R_{\alpha\beta\rho\sigma} R^{\alpha\beta\rho\sigma} g_{\mu\nu}-2 R_{(\mu \mid \beta \gamma \delta} R_{\mid \nu)}{}^{\beta \gamma \delta} + 4 \nabla^{\alpha} \nabla^{\beta} R_{\alpha(\mu \nu) \beta} \right) \notag\\
    &+\lambda a_{2}\left(\frac{1}{2} g_{\mu \nu} R_{\alpha \beta \gamma \delta} F^{\alpha \beta} F^{\gamma \delta} + 3 R^{\alpha \beta \gamma}{ }_{(\mu} F_{\nu) \gamma} F_{\alpha \beta}+2 \nabla^{\alpha} \nabla^{\beta}\left(F_{\alpha(\mu} F_{\nu) \beta}\right)\right) \notag\\
    &+\lambda a_{3}\left(\frac{1}{2}\left(F^{2}\right)^{2} g_{\mu \nu} - 4 F^{2} F_{\mu \gamma} F_{\nu}{ }^{\gamma}\right) + \lambda a_{4}\left(\frac{1}{2} F^{4} g_{\mu \nu}-4 F_{\mu \gamma} {F^\gamma}_\delta {F^\delta}_\lambda {F^\lambda}_\nu \right).
\end{align}
Similarly, by varying \eqref{action} with respect to the gauge field we get,
\begin{align}
\label{Max}
&\nabla_{\beta} F^{\beta \alpha} = - 4 \lambda a_{2} \nabla_{\beta}\left(R^{\alpha \beta \gamma \delta} F_{\gamma \delta}\right) + 8 \lambda a_{3} \nabla_{\beta}\left(F^{2} F^{\beta \alpha} \right)+8 \lambda a_{4} \nabla^{\lambda}\left(F^{\alpha}{}_{\beta} {F^\beta}_{\gamma} F^{\gamma}{ }_{\lambda}\right).
\end{align}
The generic form of the solutions to equations (\ref{EH}) and (\ref{Max}) for a static spherically symmetric spacetime is given by the following metric
\begin{eqnarray}\label{gensol}
ds^2 &=& -g(r) dt^2 +\frac{dr^2}{f(r)}+r^2\left[d\psi^2+\sin^2\psi(d\theta^2+\sin^2\theta d\phi^2)\right].
\end{eqnarray}
We have solved the equations treating higher derivative corrections perturbatively, i.e. treating $\lambda$ as a small perturbative parameter. We focus on the electrically charged solutions only. At leading order, the metric and gauge field solutions are given by
\begin{equation}
f(r)=g(r)= 1-\frac{m}{r^{2}}+\frac{q^2}{4 r^{4}}+\frac{r^2}{L^2},
\end{equation}
\begin{eqnarray}
A_t(r) = -q\frac{\sqrt{3}}{2}\left(\frac{1}{r^2} - \frac{1}{r_h^{2}} \right),
\end{eqnarray}
where $m$, $q$ are parameters at these stage and $r_h$ denotes the outer (event) horizon, which is the largest real root of the lapse function. The boundary condition imposed to fix the integration constant during the computation of the solution requires $A_t$ to vanish on the outer horizon. At first order in $\lambda$, the metric solutions take the following form \cite{Cremonini:2019wdk}
\begin{eqnarray}\label{mgsol}
f(r) &=& \Big(1-\frac{m}{r^{2}}+\frac{q^2}{4 r^{4}}+\frac{r^2}{L^2}\Big) + \lambda f_1(r)\notag\\
g(r) &=& (1+ l(r))f(r)
\end{eqnarray}
In a similar manner the solution for gauge field is,
\begin{equation}
    A_t(r) = -q\frac{\sqrt{3}}{2}\left(\frac{1}{r^{2}} - \frac{1}{r_h^{2}} \right)+\lambda A_1(r)\label{Atsol}
\end{equation}
where 
\begin{eqnarray}
  l(r) &=& 2\lambda \Big(\frac{13}{3} a_1+4 a_2 \Big) \frac{q^2 }{r^{6}},\\
f_1(r) &=& \Bigg( a_1 \Bigg( -\frac{191q^4 }{96r^{10}}- \frac{16q^2}{r^{6}} -\frac{65 q^2}{3 L^2 r^{4}} +\frac{31m q^2}{3r^{8}} +\frac{2m^2}{r^{6}}+\frac{2 r^2}{3L^4}\Bigg)\nonumber\\&&+a_2 \Bigg(-\frac{9q^4 }{4r^{10}}  -\frac{12 q^2}{r^{6}} +\frac{10mq^2}{r^{8}}- \frac{14 q^2}{L^2 r^{4}}\Bigg)-(2a_3+a_4)\frac{3q^4}{4r^{10}}\Bigg),\\ A_1(r)=&&-\frac{13 a_1 q^3}{8 \sqrt{3}}\left(\frac{1}{r^8}-\frac{1}{r_h^8}\right)+\frac{2 \sqrt{3} a_2 \left(q^3-2 m q r^2\right)}{r^8}-\frac{2 \sqrt{3} a_2 \left(q^3-2 m q r_h^2\right)}{r_h^8}\nonumber\\&&+3 \sqrt{3} \left(2 a_3+a_4\right) q^3\left(\frac{1}{r^8}-\frac{1}{r_h^8}\right).
\end{eqnarray}
While evaluating first order corrections to the metric and gauge field in $\lambda$, there arise a couple of integration constants. One of them is fixed in such a way that the gauge field vanishes at the outer horizon. We set the rest of the constants to zero, such that, at first order, AdS length has been modified.  %For the metric, one can check that the AdS length has been modified in the presence of higher derivative correction.
The modified AdS length is $L_{\text{eff}}=L\left(1+\frac{2\lambda a_1}{3L^2}\right)^{-1/2}$.

Next we will study the correction to the physical thermodynamic quantities like mass, temperature and entropy of the black hole due to the higher derivative interaction terms. The total mass of the black hole is,
\begin{align}
    M = & \frac{3 \Omega_3}{16 \pi G} ( m + \Delta m),
\end{align}
where $m$ be the uncorrected mass parameter corresponding to the black hole and $ \Delta m $ is the shift in mass due to the four-derivative correction. We compute these parameters by solving the equation $ f(r_h) = 0 $ perturbatively,
\begin{align}
m & = r_h^2 \left( 1 + \frac{r_h^2}{L^2} + \frac{q^2}{4 r_h^4} \right), \nonumber \\
\Delta m & = \lambda \left( a_1 \left( \frac{8 r_h^4}{3 L^4} - \frac{31 q^2}{3 L^2 r_h^2} + \frac{4 r_h^2}{L^2} + \frac{23 q^4}{32 r_h^8} - \frac{14 q^2}{3 r_h^4}+2 \right) \right. \notag\\
& \left. + a_2 \left(-\frac{4 q^2}{L^2 r_h^2} + \frac{q^4}{4 r_h^8} - \frac{2 q^2}{r_h^4}\right) - \frac{3 a_3 q^4}{2 r_h^8} - \frac{3 a_4 q^4}{4 r_h^8}\right).
\end{align}
The temperature of the black hole can be calculated using
\begin{equation}
    T=\frac{1}{4\pi} \sqrt{g'(r)f'(r)} \bigg|_{r_h},
\end{equation}
which gives,
\begin{align}
T = & \frac{r_h}{\pi  L^2} + \frac{1}{2 \pi r_h} - \frac{q^2}{8 \pi r_h^5} \notag\\& +\lambda \left[a_1 \left(-\frac{4 r_h}{3 \pi  L^4}-\frac{4 q^2}{3 \pi  L^2 r_h^5}-\frac{4}{\pi  L^2 r_h}-\frac{9 q^4}{16 \pi  r_h^{11}}+\frac{5 q^2}{3 \pi  r_h^7}-\frac{2}{\pi r_h^3}\right)\right.\notag\\&
  \left. +a_2 \left(-\frac{4 q^2}{\pi  L^2 r_h^5}+\frac{q^4}{4 \pi r_h^{11}}-\frac{q^2}{\pi r_h^7}\right)+\frac{3 q^4 (2 a_3+a_4)}{2 \pi r_h^{11}}\right].
\end{align}
In higher derivative gravity, the black hole entropy is calculated using the Wald entropy formula \cite{Wald:1993nt,Jacobson:1993vj}
\begin{eqnarray}\label{entropydef}
\mathcal{S} = 2\pi\int_{\partial H} d^3\Omega\; \sqrt{\tilde{h}} \; \frac{\partial {\cal L}}{\partial R^{\alpha\beta\gamma\delta}} \epsilon^{\alpha\beta}\epsilon^{\gamma\delta},
\end{eqnarray}
where $\epsilon^{\mu\nu}$ is the binormal and $\tilde{h}$ is the induced metric on the horizon. The term  $\frac{\partial {\cal L}}{\partial R^{\alpha\beta\gamma\delta}} \epsilon^{\alpha\beta}\epsilon^{\gamma\delta} $ is known as the Wald entropy density of the black hole.

In general, the binormal $\epsilon_{\mu\nu}$ is defined as $\epsilon_{\mu\nu}=\xi_\mu \eta_\nu - \xi_\nu \eta_{\mu}$ with normalization condition given by $ \epsilon_{\mu\nu} \epsilon^{\mu\nu} = -2$ \cite{Dutta:2006vs,Jiang:2018sqj}. Here $\xi$ and $\eta$ are null vectors on the horizon at the bifurcation point with $ \xi. \eta=1$. These null vectors can be taken such as
\begin{eqnarray}
\xi &=& \frac{\partial}{\partial t},\\
\eta &=&-{g(r)}^{-1} \frac{\partial}{\partial t}-\left(1-\frac{l(r)}{2}\right)\frac{\partial}{\partial r}.\label{binormal}
\end{eqnarray}
Thus the non-trivial components of the binormals are,
\begin{equation}
    \epsilon_{tr}=-\epsilon_{rt}=\left(1-\frac{l(r)}{2}\right).
\end{equation}
 For the action \eqref{action}, the Wald entropy tensor will take the form,
\begin{align}\label{waldendens}
   \Psi_{\alpha\beta\gamma\delta}= \frac{\partial {\cal L}}{\partial R^{\alpha\beta\gamma\delta}} & = \frac{1}{16\pi G}\left(\frac{1}{2}( g_{\alpha\gamma}g_{\beta\delta}- g_{\alpha\delta}g_{\beta\gamma}) + 2\lambda a_1 R_{\alpha\beta\gamma\delta} + \lambda a_2 F_{\alpha\beta}F_{\gamma\delta}\right).
\end{align}
Contracting the above tensor with the binormals, \eqref{binormal} we find
\begin{eqnarray}\label{waldendens1}
\Psi&=&\frac{\partial {\cal L}}{\partial R^{\alpha\beta\gamma\delta}} \epsilon^{\alpha\beta}\epsilon^{\gamma\delta} = \frac{2}{16\pi G}[1-4\lambda a_1 R_{trtr}-2 \lambda a_2 F_{tr}F_{tr}]\nonumber\\
  &=& \frac{1}{16\pi G}\left(2 - \lambda a_1 \left( \frac{8}{L^2} + \frac{20 Q^2}{r^6} - \frac{24 m}{r^4}\right) - 12 \lambda a_2 \frac{Q^2}{r^6}\right)
\end{eqnarray}
where the black hole charge $Q$ is defined using the Gauss' law as,
\begin{equation}
    Q = \frac{1}{16\pi G}\int_{r\rightarrow\infty} \ast F = \frac{\sqrt{3} \Omega_3 q}{16 {\pi G}}.
\end{equation}
 $\ast F$ is known as the Hodge dual of the field strength $F_{\alpha\beta}$.
Hence the Wald entropy for the black hole is
\begin{eqnarray}
\mathcal{S} = \frac{ \Omega_3 {r_h}^3}{4 G} + \frac{\lambda \Omega_3}{G} \left( - a_1 \left( \frac{r_h^3}{L^2} + \frac{5 Q^2}{3 r_h^3} - \frac{3 m}{r_h}\right) -  \frac{3 a_2 Q^2}{2 r_h^3}\right).
\end{eqnarray}
We compute the expression for chemical potential $\mu$ through the first law of thermodynamics,
\begin{equation}
    dM-TdS-\mu dQ=0
\end{equation}
The chemical potential is conjugate to electric charge and defined as the potential difference between the horizon and spatial infinity,
\begin{equation}\label{mu}
    \mu(r_h)= A_t(r\rightarrow \infty) - A_t(r\rightarrow r_h).
\end{equation}
The boundary condition implies that the second term on the r.h.s of \eqref{mu} vanishes. Using the solution for $A_t$ given in \eqref{mgsol} we find

\begin{align}
    \mu(r_h)= \frac{4G Q}{\pi r_h^2} + \lambda \Bigg(a_1 \frac{832 G^3 Q^3}{9\pi^3 r_h^8} + a_2\frac{256G^2 Q}{3\pi^3 r_h^8} \Big(M \pi r_h^2 -4G Q^2\Big)\notag\\
    - (2a_3+a_4)\frac{512 G^3 Q^3}{\pi^{3} r_h^8} \Bigg),
\end{align}
where we have used the expressions of $m$ and $q$ in terms of the mass $M$ and charge $Q$ of the black hole.
\section{Computation of holographic complexity}\label{sec2}

The $U(1)$ gauge field in the AdS bulk dual to a global $U(1)$ symmetry generating conserved current at the boundary CFT. The charged eternal black hole \eqref{gensol} we considered in section \ref{sec1} is dual to two entangled CFTs at the left and right boundaries in the charged thermofield double state (TFD). The eternal black hole is symmetric under $t\rightarrow -t$, i.e., the gravitational solution \eqref{gensol} is invariant under $t_L\rightarrow t_L + \Delta t, t_R \rightarrow t_R -\Delta t $. This corresponds to boost symmetry in the Kruskal coordinate and the invariance of the dual TFD state under an evolution with Hamiltonian $H = H_L-H_R$. However, we are interested in studying the complexity growth for $t>0$, thus will consider the evolution of the time-dependent TFD state,
\begin{align}
    |\Psi((t_L, t_R))\rangle =\frac{1}{\sqrt{Z(\beta)}}\sum_{n, \sigma}e^{-\beta(E_{n}-\mu Q_{\beta})/2}e
^{-iE_n(t_L+t_R)}|n,-\sigma\rangle_L |n,\sigma\rangle_R\label{TFD}
\end{align}
under the Hamiltonian $H_L+H_R$. In \eqref{TFD} $\beta$ is the inverse temperature, $E_n$ is the energy eigenvalue and $Q_{\beta}$ is the eigenvalue under the charge operator for the total system. This TFD state involves a chemical potential $\mu$ which distinguishes the boundary states by their U(1) charges. We are interested to compute the growth of holographic complexity of the TFD state, following the CA conjecture. This conjecture tells us that the complexity of boundary states is proportional to the gravitational action evaluated in the Wheeler-De Witt (WDW) patch, i.e. the codimension 0 bulk region bounded by the null surfaces anchored at left and right boundary at a chosen time. Therefore,
\begin{equation}
C_A =\frac{I_{\text{WDW}}}{\pi\hbar}
\end{equation}

where the total action $I_{\text{WDW}}$ can be written as,
\begin{eqnarray}
I_{\text{WDW}} = I_{\text{bulk}}+I_{\text{boundary}}+I_{\text{joint}}+I_{\text{ct}}
\end{eqnarray}
$I_{\text{bulk}}$ denote contribution from the bulk action \eqref{action} on the patch. $I_{\text{boundary}}$ comes from the boundary action \eqref{bdryaction} for smooth timelike and
spacelike segments of the boundaries, whereas $I_{\text{joint}}$ is evaluated at the intersection of spacelike/timelike/null boundaries. $I_{\text{ct}}$ has been added to maintain the the full action invariant under reparametrization of null surfaces.

\begin{figure}[!htb]
\centering
\begin{minipage}{0.3\textwidth}
\centering
\includegraphics[scale=0.3]{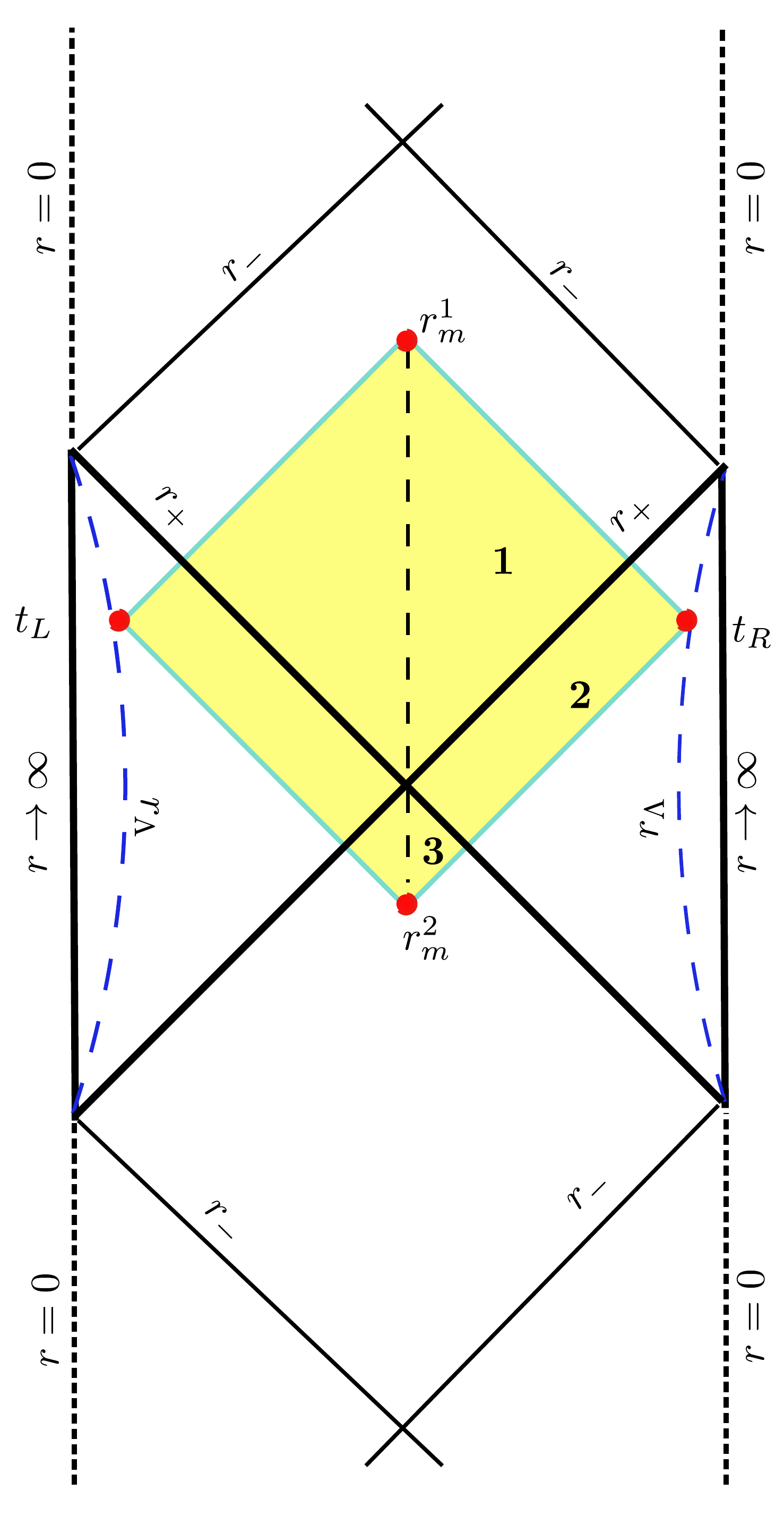}
\label{fig:1}
\end{minipage}
\begin{minipage}{0.4\textwidth}
\centering
\includegraphics[scale=0.3]{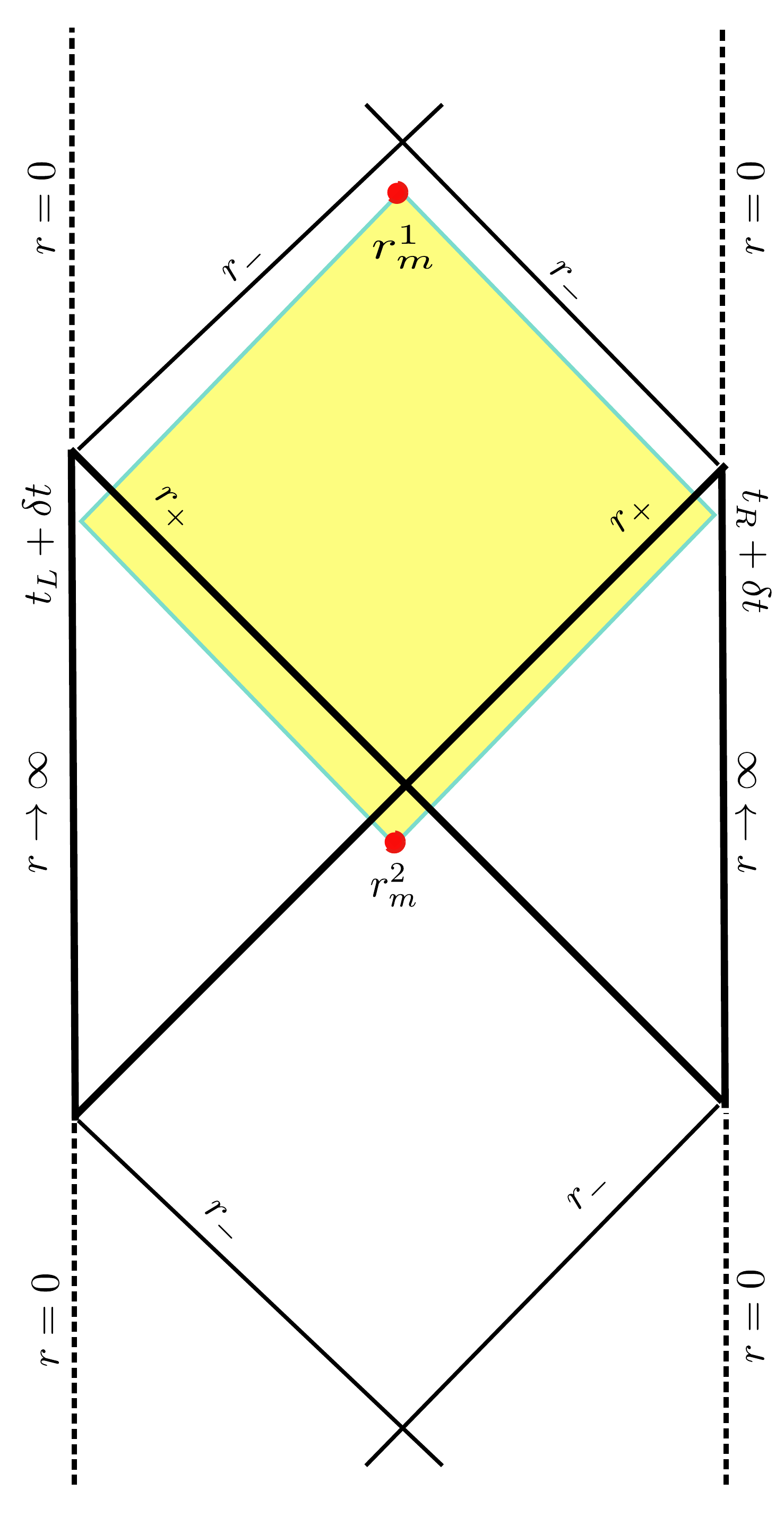}
\label{fig:2}
\end{minipage}
\caption{Penrose diagram for the charged black hole, Left: The WDW patch at boundary time $t_L=t_R=t/2$; Right: The evolved WDW patch after time interval $\delta t$}
\end{figure}
In general, the charged black hole \eqref{gensol} has two horizons. We denote the event horizon as $r_{h}=r_{+}$ and the inner horizon as $r_{-}$. We can consider the WDW patch as the Yellow region in Figure 1. The WDW patch is bounded by the null surfaces originated at asymptotic boundaries at time $t_L$ and $t_R$. Complexity growth rate of the boundary TFD state at a specific $t_L$ and $t_R$ are thus proportional to the change of gravitational action with the boundary time, evaluated on WDW patch. We can set without loss of generality $t_L=t_R=\frac{t}{2}$. To avoid divergences at the timelike boundary of the AdS spacetime we introduce an UV cut-off $r_{\Lambda}$. 
The WDW patch has four vertices intersecting two null boundaries. Two of which are at the cutoff surface at time $t_L$ and $t_R$. We denote the radial position of the top and bottom vertices of the WDW patch by $r_m^1$ and $r_m^2$. 
We define the Eddington-Finkelstein coordinates as
\begin{align}
    u=t-r^{*}(r),\quad v=t+r^{*}(r)
\end{align}
where $r^{*}(r)$ is the tortoise coordinate, 
\begin{equation}
     r^{*}(r)=-\int_r^\infty ~dr\sqrt{\frac{g_{rr}}{g_{tt}}}=-\int_r^\infty~dr\sqrt{\frac{1}{f(r)g(r)}}
\end{equation}
We begin with the TFD state at $t=0$, which fixes $r_m^1$ and $r_m^2$ in terms of boundary time and UV cut off according to,
\begin{equation} 
t_L+ r^*_{\Lambda} - r^*(r_m^{1})=0,
\qquad
t_R-r^*_{\Lambda}+r^*(r_m^{2})=0,\label{rm}
\end{equation}
By differentiating Eq. \eqref{rm}, we find
\begin{equation} 
\frac{d r_m^{1}}{d t} = \frac{\sqrt{f(r_m^{1})g(r_m^{1})}}{2},
\qquad
\frac{d r_m^{2}}{d t} = - \frac{\sqrt{f(r_m^{2})g(r_m^{2})}}{2}.\label{rmt}
\end{equation}
In the right panel of Figure 1, the WDW patch has moved upwards with respect to advances $\delta t$ in boundary time. In the late time limit, the future joint at $r_m^{1}$ touches the inner horizon $r_{-}$ and the past joint at $r_m^{2}$ touches the outer horizon $r_{+}$. Therefore the time evolution of the action is  determined by the evolution of $r_m^1$ and past $r_m^2$ following \eqref{rmt}. In the following subsections we will compute the various contributions to the gravitational action on the WDW patch. Due to the symmetry of the Penrose diagram, we have divided the WDW patch across the dashed line %, which is $t=0$ surface
in Figure 1. We denote three regions as 1, 2 and 3 on the right part of WDW patch. Region 1 is behind the future horizon $r^+$, region 2 is outside of both the horizons touching spatial infinity, whereas region 3 is behind the past horizon. We will calculate the gravitational action in these three regions and multiply with the factor of two to obtain the contribution from the full patch.

\subsection{Bulk contributions }
In this subsection, we compute the contribution from the bulk action on the WDW patch following \cite{Carmi:2017jqz}, 
\begin{equation}
    I_{bulk}=\Omega_3 \int dt\int dr \; r^3 \sqrt{(1+ l(r) )} \,  \mathcal{L}_{OS}
\end{equation}
where $\mathcal{L}_{OS}$ is the on-shell Lagrangian which can be computed by plugging the metric solution \eqref{mgsol} in the bulk Lagrangian of \eqref{action},
\begin{eqnarray}
&&\mathcal{L}_{OS} =\frac{1}{16 \pi G} \Bigg(-\frac{8}{L^2}+\frac{q^2}{r^6}+\lambda \Big[a_1\Big(\frac{80}{3 L^4}-\frac{20 q^2}{3 L^2 r^6}+\frac{48 m^2}{r^8}-\frac{36 m q^2}{r^{10}}+\frac{79 q^4}{6 r^{12}}-\frac{16 q^2}{r^8}\Big) \nonumber\\&& +a_2 \Big( -\frac{8 q^2}{l^2 r^6}+\frac{72 m q^2}{r^{10}}-\frac{24 q^4}{r^{12}}-\frac{48 q^2}{r^8}\Big)-(2a_3+a_4)\frac{12q^4}{r^{12}}\Big]\Bigg).
\end{eqnarray}
The total on-shell bulk action can be written as a sum of contributions from the three parts of the WDW patch as follows
\begin{align}
I^{\rm I}_{\rm bulk} &=2\Omega_3 \,\int^{r_+}_{r^1_m}  dr\, r^3\,\sqrt{(1+ l(r) )} \mathcal{L}_{OS}(r) \left(\frac{t}{2}+r^*_{\Lambda}-r^*(r)\right) ,\nn\\
I^{\rm II}_{\rm bulk} &= 4\Omega_3 \,\int^{r_{\Lambda}}_{r_+}  dr\, r^3\, \sqrt{(1+ l(r) )} \mathcal{L}_{OS}(r)(r^*_{\Lambda}-r^*(r) ) ,\nn\\
I^{\rm III}_{\rm bulk} &=2\Omega_3 \,\int^{r_+}_{r^2_m}  dr\, r^3\, \sqrt{(1+ l(r) )} \mathcal{L}_{OS}(r) (-\frac{t}{2}+r^*_{\Lambda}-r^*(r)). 
\end{align}

where $\Omega_3=\frac{2\pi^2}{\Gamma(2)}=2\pi^2$. An additional multiplication of a factor of $2$ is required due to the two symmetric sides of the Penrose diagram. The time derivative of bulk on-shell action is, 
\begin{align}
\frac{d I_{\rm bulk}}{d t} &=
\left(\Omega_3 \int_{r_m^{1}}^{r_+}  dr~ r^3 \sqrt{(1+ l(r) )}\mathcal{L}_{OS}(r) -\Omega_3 \int_{r_m^{2}}^{r_+} dr ~r^3 \sqrt{(1+ l(r) )} \mathcal{L}_{OS}(r) \right)\nonumber\\ 
&= \Omega_3 \int_{r_m^{1}}^{r_m^{2}} dr ~r^3 \sqrt{(1+ l(r) )} \mathcal{L}_{OS}(r)\nonumber\\&
=  \frac{\Omega_3}{16\pi G}\Bigg[-\frac{2 r^4}{L^2}-\frac{q^2}{2 r^2}+\lambda \Bigg\{a_1\left( \frac{20 r^4}{3 L^4}+\frac{62 q^2}{3 L^2 r^2}+\frac{4 \left(q^2-3 m^2\right)}{r^4}+\frac{6 m q^2}{r^6}-\frac{35 q^4}{16 r^8} \right)\nonumber\\&+a_2 \left( \frac{20 q^2}{L^2 r^2}-\frac{12 m q^2}{r^6}+\frac{5 q^4}{2 r^8}+\frac{12 q^2}{r^4} \right)+(2a_3+a_4)\frac{3 q^4}{2 r^8}\Bigg\}\Bigg]_{r_m^1}^{r_m^2}\label{bulk/t}
\end{align}
 At late times, $r^{1}_m\to r_{- }$ and $r^2_m\to r_+ $. Therefore at late times \eqref{bulk/t} gives,
\begin{eqnarray}
\frac{d I_{\rm bulk}}{d t}\bigg|_{t\rightarrow\infty}=\Bigg[\frac{2}{3}M-TS-\mu Q + \lambda  \left(a_1 \left(\frac{2 \pi  r^4}{L^4}+\frac{32 Q^2+6 \pi ^2 r^4}{3 \pi  L^2 r^2}\right)-\frac{32 a_2 Q^2}{\pi  L^2 r^2}\right)\Bigg]_{r_-}^{r_+}\nonumber\\
\end{eqnarray}
where we have rewritten the expression in terms of mass, charge, temperature and entropy for the complete black hole solution.

%%%%%%%%%%%%%%%%%%%%%%%%%%%%%%%%%
\subsection{Boundary contributions}
The WDW patch has four null boundary surfaces. Henceforth we have to consider the contribution of the null boundary terms to the gravitational action. The null surface can be foliated by an outward null normal to the hypersurface $k_{\alpha}$ and an auxiliary null vector transverse to the hypersurface $N^{\alpha}$ satisfying $k_{\alpha}N^{\alpha}=-1$. We can write the induced metric on the null surface in terms of these quantities as, 
\begin{equation}
    \sigma_{\alpha\beta}=g_{\alpha\beta}+k_{\alpha} N_{\beta}+N_{\alpha} k_{\beta}\label{nm}
\end{equation}
The null hypersurface spanned by $\varsigma, \vartheta^A$ can be parametrized by $x^{\alpha}=x^{\alpha}(\varsigma, {\vartheta}^A)$ where $\varsigma$ is a parameter along the null generator and $\vartheta^A$ is constant on each null generator. One can define vectors $k^{\alpha}, e^{\alpha}_A$ tangent to the hypersurface,
\begin{align}
    k^{\alpha}=\frac{\partial x^{\alpha}}{\partial \varsigma}, \quad  e^{\alpha}_A=\frac{\partial x^{\alpha}}{\partial \vartheta^A}
\end{align}
The induced metric $\sigma_{\alpha\beta}$ is orthogonal to both the null normal $k^{\alpha}$ and auxiliary null vector $N^{\alpha}$. From \eqref{nm} one can easily see that the projector is as following,
\begin{equation}
   \sigma^{\beta}_{\nu}= \sigma^{\beta\mu}\sigma_{\mu\nu}=\delta^{\beta}_{\nu}+k_{\nu}N^{\beta}+k^{\beta}N_{\nu}
\end{equation}
One can write the induced degenerate spatial metric on the null hypersurface as,
\begin{align}
    \sigma^{\alpha\beta}=e^{\alpha}_Ae^{\beta}_B \sigma^{AB}
\end{align}
and the inverse metric as
\begin{equation}
    g^{\alpha\beta}=e^{\alpha}_A e^{\beta}_B \sigma^{AB}-(k^{\alpha} N^{\beta}+N^{\alpha} k^{\beta})
\end{equation}
The acceleration of the null vector $k^{\beta}$ satisfies 
\begin{equation}
    k^{\alpha}\nabla_{\alpha}k^{\beta}=\kappa k^{\beta}
\end{equation}
where $\kappa(\varsigma,\vartheta^A)$ is known as the non-affinity coefficient which measures the failure of $\varsigma$ to be an affine parameter. We need to introduce a boundary term in the gravitational action for a well behaved variation of the bulk action on the null surface \cite{Jiang:2018sqj,Jiang:2020spf},
\begin{align}
    \int dS~d\lambda~ [2k_{\rho}\Psi_{\alpha}{}^{\beta\rho\delta}\delta \Gamma^{\alpha}_{\beta\delta}+2k_{\rho}\delta g_{\beta\delta}\nabla_{\alpha}\Psi^{\alpha\beta\rho\delta}]
\end{align}
where $\Psi_{\alpha\beta\gamma\delta}$ is the Wald entropy tensor introduced earlier in \eqref{waldendens}. After some manipulations, we can write the contribution to the action from the null surface as \cite{Jiang:2018sqj},
\begin{align}\label{boundary}
    I_{\text{boundary}}=-\int dS d\varsigma\Psi\kappa 
\end{align}
Here $\Psi= 4\Psi_{\alpha\beta\gamma\delta}k^{\alpha}N^{\beta}k^{\gamma}N^{\delta}=\Psi_{\alpha\beta\gamma\delta}\epsilon^{\alpha\beta}\epsilon^{\gamma\delta}$ \cite{Jiang:2018sqj}. However \eqref{boundary} vanishes if one chooses an affine parametrization of the null generators implying $\kappa=0$.
\subsection{Joint contributions }
The WDW patch has four joints at the intersection of two null boundary surfaces. Two joints are located at the UV cut-off $r=r_\Lambda$. Others are intersecting points of past and future null boundaries of the WDW patch situated at $r=r_m^2$ and $r=r_m^1$ respectively. It can be checked that, the joint terms evaluated at the UV cut-off surface is time independent and do not contribute to the time growth rate of the complexity. The joint term computed at $r_m^1$ and $r_m^2$ only take part non-trivially to the time derivative of the holographic complexity since the position of $r_m^1$ and $r_m^2$ changes with time according to \eqref{rmt} as the WDW patch evolves.\\~~~\\
At the intersection of two null boundaries for the higher derivative gravity theories, the joint term is given by \cite{Cano:2018aqi, Cano:2018ckq, Jiang:2018sqj}
\iffalse
\begin{equation}
    I_{\text{Joint}}=\frac{1}{2\pi}\int d\sigma a~S
\end{equation}
where
\begin{equation}
S_{JM}=-2\pi \int_{\partial H} \frac{\delta I}{\delta R_{ABCD}}\epsilon_{AB}\epsilon_{CD}\sqrt{h}d^{D-2}\Omega;\quad {\partial H}=\text{horizon cross section}
\end{equation}
and 
\fi
\begin{equation}
    I_{\text{Joint}}=\int dS ~a \Psi\label{nulljoint}
\end{equation}
where the Wald entropy density $\Psi$ is given by \eqref{waldendens1} and $a(r)$ is defined as \cite{Lehner:2016vdi},
\begin{equation}
    a(r)=\eta\ln\left(\frac{|k_L.k_R|}{2}\right)%=-\eta\log\left(\frac{|g(r)|}{\alpha^2}\right)
    \label{a}
\end{equation}
where $\eta$ takes the value of 1 or $-1$ depending on the position of joints. For joints at the UV cut off we have $\eta=-1$ and for the past, future joints $\eta=1$. $k_L$ and $k_R$ are null normals to both the segments intersecting at the joint. We use the following null normal vectors,  
\begin{align}
 &k_{R} = \alpha (-dt +dr^{*})= \alpha \left(-dt+\frac{dr}{\sqrt{f(r)g(r)}}\right)\,\notag\\&
 k_{L} = \alpha (dt +dr^{*})= \alpha \left(dt+\frac{dr}{\sqrt{f(r)g(r)}}\right)\,,
\end{align}
which are outward-directed towards right and left respectively. These null normals satisfy
\begin{equation}
    k_L . k_R=\frac{2\alpha^2}{g(r)}
\end{equation}
The next step is evaluating the joint contributions coming from the meeting points, $r=r_m^1$ and $r_m^2$. Using \eqref{nulljoint} and \eqref{entropydef} we can obtain the contribution from both the joint terms at $r_m^1$ and $r_m^2$ as,
\begin{equation}
    I_{joint}=\frac{1}{2\pi}\left(\mathcal{S}(r_m^1)a(r_m^1)+\mathcal{S}(r_m^2)a(r_m^2)\right)
\end{equation}
where $\mathcal{S}$ is evaluated at each respective null surfaces.

%%%%%%%%%%%%%%%%%%%%%%%%%%%%
We will consider the growth of $I_{\text{joint}}$ in time. At the meeting point at $r_m^1$, it can be checked that, 
\begin{eqnarray}\label{jen}
\frac{d(a~\mathcal{S})}{dt}\bigg|_{r_m^1}&=& ~\left(\frac{da}{dr}~\mathcal{S}+a\frac{d\mathcal{S}}{dr}\right)\bigg|_{r_m^1}\frac{dr_m^1}{dt}
\end{eqnarray}
Using the definition of $a$ from \eqref{a} and the relation $\frac{dr_m^1}{dt}=\frac{\sqrt{f(r_m^1)g(r_m^1)}}{2}$, equation \eqref{jen} can be expressed as
\begin{eqnarray}\label{jen1}
\frac{d(a~\mathcal{S})}{dt}\bigg|_{r_m^1}&=& -\frac{\sqrt{f(r_m^1)g(r_m^1)}}{2}\left(\frac{g'(r_m^1)}{g(r_m^1)}\mathcal{S}(r_m^1)+\log\left(\frac{|g(r_m^1)|}{\alpha^2}\right)\frac{d\mathcal{S}}{dr}\bigg|_{r_m^1}\right)
\end{eqnarray}
Similarly, using the relation $\frac{dr_m^2}{dt}=-\frac{\sqrt{f(r_m^2)g(r_m^2)}}{2}$, we find
\begin{eqnarray}\label{jen11}
\frac{d(a~\mathcal{S})}{dt}\bigg|_{r_m^2}&=& \frac{\sqrt{f(r_m^2)g(r_m^2)}}{2}\left(\frac{g'(r_m^2)}{g(r_m^2)}\mathcal{S}(r_m^2)+\log\left(\frac{|g(r_m^2)|}{\alpha^2}\right)\frac{d\mathcal{S}}{dr}\bigg|_{r_m^2}\right)
\end{eqnarray}
Hence we can write the time growth rate of the joint term as
\begin{eqnarray}
 \frac{dI_{joint}}{dt}&&=\frac{\sqrt{f(r_m^2)g(r_m^2)}}{4\pi}\left(\frac{g'(r_m^2)}{g(r_m^2)}\mathcal{S}(r_m^2)+\log\left(\frac{|g(r_m^2)|}{\alpha^2}\right)\frac{d\mathcal{S}}{dr}\bigg|_{r_m^2}\right)\nonumber\\
&&-\frac{\sqrt{f(r_m^1)g(r_m^1)}}{4\pi}\left(\frac{g'(r_m^1)}{g(r_m^1)}\mathcal{S}(r_m^1)+\log\left(\frac{|g(r_m^1)|}{\alpha^2}\right)\frac{d\mathcal{S}}{dr}\bigg|_{r_m^1}\right)\label{joint}
\end{eqnarray}
At late time, when $r_m^1\rightarrow r_-$ and $r_m^2\rightarrow r_+$. In this limit, the time derivative of the joint term becomes
\begin{eqnarray}
\frac{dI_{joint}}{dt}\bigg|_{t\rightarrow\infty}&=& T(r_+){\mathcal S}(r_+)-T(r_-){\mathcal S}(r_-)
\end{eqnarray}
where we have manipulated the first term in \eqref{joint} as,
\begin{align}
 \frac{\sqrt{f(r_+)g(r_+)}}{4\pi}\frac{g'(r_+)}{g(r_+)}&=T(r_+)\sqrt{\frac{f(r_+)g'(r_+)}{g(r_+)f'(r_+)}}
 \notag\\&=T(r_+)\sqrt{1+l'(r_+)\frac{f(r_+)}{f'(r_+)}}\nonumber\\
 &=T(r^+)
\end{align}
Similar calculation holds true at $r=r_-$. We have used the relation \eqref{mgsol} between $f(r)$ and $g(r)$ in the above calculation. 
%%%%%%%%%%%%%%%%%%%%%%%%%

%%%%%%%%%%%%%%%%%%%%%%%%%%%%%%%

\subsection{Counterterm contribution}
Evaluation of the surface term at the null boundary and also the joint term at meeting point with a null-boundary depends on the choices of parametrization of the null generators. Different choices of parametrization may lead to different results. This ambivalence property can be avoided by adding a proper counter term to accommodate the contribution from the null boundary $\mathcal{N}$ \cite{Jiang:2018sqj},
\begin{eqnarray}
I_{ct} &=& -\int_{\mathcal{N}} d\varsigma dS ~\nabla_{\alpha}\left(k^{\alpha}\Psi\right)\log\left(l_{ct} \nabla_{\alpha} k^{\alpha}\right)
\end{eqnarray}
with $\Psi$ being the Wald entropy density, $k^{\alpha}$ is the null normal and $l_{ct}$ is an arbitrary length scale. We can write $\nabla_\alpha k^\alpha=\frac{\partial_\varsigma \sqrt{\sigma}}{\sqrt{\sigma}}$ and $\nabla_{\alpha}\left(k^{\alpha}\Psi\right)=\frac{\partial_\varsigma (\sqrt{\sigma}\Psi)}{\sqrt{\sigma}}$ where $\sigma$ is the induced metric on the three-sphere. Affinely parametrizing  $\varsigma$ on each null generator, we can write it as $\varsigma=\frac{r}{\rho}$, where $\rho$ is an arbitrary constant. In effect, the counterterm is
\begin{equation}
    I_{ct} = -2\pi^2\left(\int_{r_m^1}^{r_{max}} dr {\Phi^{'}}(r)\ln\left(\frac{3l_{ct}\rho}{r}\right)+\int_{r_m^2}^{r_{max}} dr {\Phi^{'}}(r)\ln\left(\frac{3l_{ct}\rho}{r}\right)\right)
\end{equation}
where $\Phi(r)=\frac{2}{16\pi G}(1-\kappa(4a_1 R_{trtr}+2a_2 F_{tr}F_{tr}))r^3$. 
Then we find that
\begin{equation}
\frac{dI_{ct}}{dt}=-\pi^2
\left[\sqrt{f(r)g(r)} {\Phi^{'}}(r)\ln\left(\frac{3l_{ct}\rho}{r}\right)
\right]
_{r_m^1}^{r_m^2}
\end{equation}
As the meeting points of the null boundaries reaches the horizon at late time, the above term vanishes. Thus
\begin{equation}
   \frac{dI_{ct}}{dt}\bigg|_{t\rightarrow\infty}=0 
\end{equation}
\subsection{Maxwell boundary term}
The Maxwell boundary term in the computation of holographic complexity was introduced to retain the late time growth for purely magnetic black hole similar to the electric black hole \cite{Goto:2018iay}. This helps to restore the electromagnetic duality in four spacetime dimensions. Fixing the coefficient in this boundary term to certain values results in a vanishing contribution to the holographic complexity late time growth for the electric black hole. The variation of the surface term gives terms proportional to $\delta F_{\mu\nu}$ together with the $\delta A_{\mu}$ terms. Therefore addition of the surface term as in \eqref{gaugevariation} implies the modification of the boundary conditions. For the Maxwell case, if one chooses the coefficient $\gamma=1$, the terms proportional to $\delta A_{\mu}$ vanishes and the term proportional to $\delta F_{\mu\nu}$ can be eliminated by imposing the Neumann boundary condition. If we consider an arbitrary value of $\gamma$ it will force us to consider the mixed boundary condition for the gauge field. 

We will convert the surface term of \eqref{gaugevariation} using the Stoke's theorem and plugging in the equation of motion \eqref{Max} in it we get, 
\begin{align}\label{muQons}
    I_{\mu Q}|_{\text{on-shell}}=\frac{\gamma}{16 \pi G}\int d^5 x\sqrt{-g}\Bigg(\frac{1}{2} F^2 -\lambda \Big( 2a_2 R^{\mu\nu\alpha\beta}F
    _{\mu\nu}F_{\alpha\beta} + 4a_3 (F^2)^2 + 4a_4 F^4\Big)\Bigg)
\end{align}
By using explicit solutions \eqref{mgsol}, it can be shown that at late time \eqref{muQons} simplifies to,
\begin{eqnarray}
    \frac{d  I_{\mu Q}}{dt}\bigg|_{t\rightarrow\infty} &=& \gamma\bigg[\frac{4Q^2}{\pi r^2}+\lambda \bigg(a_1 \frac{832Q^4}{9\pi^3r^8}+a_2\bigg(\frac{32Q^2}{\pi r^4}+\frac{64 Q^2}{\pi L^2 r^2}-\frac{512 Q^4}{3\pi^3 r^8}\bigg)-\nonumber\\&&(2a_3+a_4)\frac{512Q^4}{\pi^3 r^8}\bigg)\bigg]_{r_-}^{r^+}
\end{eqnarray}
\subsection{Total late time growth rate}
Adding all the contributions, we find that the total late time growth rate of the action complexity is
\begin{eqnarray}\label{final}
   \frac{d\mathcal{C}_A}{dt}\bigg|_{t\rightarrow\infty} &=&\frac{1}{\pi}  \frac{dI_{tot}}{dt}\bigg\vert_{t\rightarrow\infty} \nonumber\\
   &=&\bigg[ \frac{2}{3} M +(\gamma-1)\mu Q+\lambda\bigg(a_1\bigg(\frac{32Q^2}{3 \pi L^2 r^2}+\frac{2\pi r^2}{L^2}\bigg(1+\frac{r^2}{L^2}\bigg)\bigg)\nonumber\\&&+a_2(\gamma-1)\frac{32Q^2}{\pi L^2 r^2}\bigg) \bigg]_{r_-}^{r^+}
\end{eqnarray}
We can see that, the nature of the total late time growth rate depends on the choices of $\gamma$. If we set $\gamma=1$, the dependence of  $\frac{d\mathcal{C}_A}{dt}|_{t\rightarrow\infty}$ on $a_2$ drops. If we also set $a_1=0$, this result is consistent with the late time complexity growth of a Reissner-Nordstrom black hole. In the case of a dyonic black hole in four dimension \cite{Goto:2018iay, Razaghian:2020bfk}, the late time complexity growth can be solely expressed in terms of the magnetic charges when $\gamma=1$. The contribution from electric charge to $\frac{d\mathcal{C}_A}{dt}|_{t\rightarrow\infty}$ drops to zero. For $\gamma=1/2$ the complexity growth depends on the combination of electric and magnetic charge on an equal footing. 

We find that the late time growth rate of the action complexity explicitly depends on the two higher derivative terms $R_{\mu\nu\alpha\beta}R^{\mu\nu\alpha\beta}$ and $R_{\mu\nu\alpha\beta}F^{\mu\nu}F^{\alpha\beta}$. This finding indicates the possibility of the violation of LLoyd's bound at late times which says that for charged black hole, $\frac{d\mathcal{C}_A}{dt}|_{t\rightarrow\infty}\leq (M-\mu Q)\vert_{r_+}- (M-\mu Q)\vert_{r_-}$. This relation fails in the presence of Maxwell boundary term with non trivial $\gamma$ even when the gravity theory contains at most two derivative terms \cite{Goto:2018iay}. A violation was also recently observed in \cite{Babaei-Aghbolagh:2021ast} for some specific momentum relaxation terms. We see that, in our case, even when $\gamma=0$ the late time complexity growth has non-trivial dependence on the higher derivative terms. %correspond to $R_{\mu\nu\alpha\beta}R^{\mu\nu\alpha\beta}$ and $R_{\mu\nu\alpha\beta}F^{\mu\nu}F^{\alpha\beta}$. 
This indicates that for our gravitational action  $\frac{d\mathcal{C}_A}{dt}|_{t\rightarrow\infty}$ may violate the Lloyd's bound for any choices of $\gamma$ depending on the respective value and sign of $a_1$ and $a_2$. As an example, by comparing with a higher derivative supergravity theory \cite{Cremonini:2008tw}, we can set $a_2=-\frac{1}{2}a_1$. In this case, we explicitly check that the $\frac{d\mathcal{C}_A}{dt}|_{t\rightarrow\infty}$ always violate Lloyd bound as long as $a_1$ is positive. This is a new finding compared to the other cases in presence of higher derivatives \cite{Razaghian:2020bfk,Ghodsi:2020qqb,Cano:2018aqi}.

\subsection{Comparison with violation of KSS bound}

One important discovery in the AdS/CFT duality was realization of the fact that the hydrodynamic properties of the black-brane horizon can be identified with the same of the boundary CFT theories at thermal equilibrium in the low frequency and vanishing momentum limit. Following the effective theory description one can expand the energy-momentum tensor in powers of the spacetime derivatives. At zeroth order we get the equations for ideal fluid. The next order derivative expansion provides us the dissipative behaviour and one can compute the diffusion coefficients like shear  viscosity, the bulk viscosity and the heat conductivity. For a large class of four dimensional CFT with Einstein gravity duals the ratio of shear viscosity $\eta$ and entropy density $s$ is a universal constant and bounded as \cite{Kovtun:2004de} ,
\begin{equation}
    \frac{\eta}{s}\geq\frac{1}{4\pi}
\end{equation}
This is known as the KSS bound. The shear viscosity of the strongly coupled field theory in this context can be computed from field theory correlators dual to the shear gravitational perturbations, using the Kubo formula. However the higher curvature corrections modify this ratio and the universal lower bound does not hold anymore \cite{Dobado:2008ri}. In \cite{Myers:2009ij} it was observed that increase in the chemical potential in the strongly coupled gauge theory due to the higher derivative interaction term in the dual gravitational theory amplify the violation of KSS bound. 

For the gravitational theory coupled to a gauge field described by the action \eqref{action}, the ratio for the dual field theory is
\begin{equation}
    \frac{\eta}{s}=\frac{1}{4\pi}\left[1-8 a_1+ 4(a_1+6 a_2)\frac{q^2}{r_{+}^6}\right]
\end{equation}
Hence the KSS bound is violated for positive value of $a_1$ and negative value of $a_2$. For the same example of a higher derivative supergravity theory \cite{Cremonini:2008tw}, with $a_2=-\frac{1}{2}a_1$ \cite{Myers:2009ij} the violation of the lower bound is explicitly realized. This violation occurred due to the consideration of $a_1 R_{\mu\nu\alpha\beta}R^{\mu\nu\alpha\beta}$ and $a_2 R_{\mu\nu\alpha\beta}F^{\mu\nu}F^{\alpha\beta}$ terms in the gravitational action. The reason for the violation of Llyod's bound in \eqref{final} is also these two higher derivative terms. In \cite{Parvizi:2022lbv} an attempt to relate the KSS bound and the Llyod's bound has been taken. Unlike the fluid/gravity correspondence an analogue gravity model is considered which relates a gravitational theory to a fluid in the bulk only on a time-like finite cut-off. By using the CV2.0 proposal both the bounds are shown to be equivalent to each other. %While we can't comment about the relation between these two bounds 
We identify that the violation of both the bounds for our case happens due to the same terms in the action.

\section{Complexity growth of higher derivative corrected JT gravity}\label{sec3}

In this section, we study the complexity growth of one dimensional CFT that is dual to a higher derivative corrected JT like theory in two dimension. The two dimensional JT like gravitational theory we are considering is constructed upon dimensional reduction of the following  four dimensional action  \cite{Banerjee:2021vjy},
\begin{eqnarray}\label{4Dbulk1}
	\hat{S}_{bulk}&& = \frac{1}{16\pi G_4} \int d^4x \sqrt{-\hat{g}} \Big( \hat{R} - 2\Lambda -\hat{F}_{AB}\hat{F}^{AB} + \lambda\alpha_1\hat{R}^2 + \lambda\alpha_2\hat{R}_{AB}\hat{R}^{AB} \nonumber\\&&+ \lambda\alpha_3\hat{R}_{ABCD}\hat{R}^{ABCD}  \Big), 
\end{eqnarray}
where $G_4$ is the four dimensional Newton's constant. $\lambda$ is a dimensionless constant that controls whether the higher derivative corrections are perturbative or non-perturbative. $\alpha_1, \alpha_2 $ and $\alpha_3$ are finite constant with the dimension of $length^2$. We will treat the higher curvature terms perturbatively thus $\lambda$ is very small. The higher derivative parts of the action can be rewritten as,
\begin{eqnarray}
	\alpha_1\hat{R}^2 + \alpha_2\hat{R}_{AB}\hat{R}^{AB} + \alpha_3\hat{R}_{ABCD}\hat{R}^{ABCD} = \tilde{\alpha}_1\hat{R}^2 + \tilde{\alpha}_2\hat{R}_{AB}\hat{R}^{AB} + \alpha_3\hat{R}_{GB}^2.
\end{eqnarray}
Where $\hat{R}_{GB}^2=\hat{R}^2-4\hat{R}_{AB}\hat{R}^{AB}+\hat{R}_{ABCD}\hat{R}^{ABCD}$ is the Gauss-Bonnet term, which is a topological term in four-dimension.
To rewrite we have redefined the coefficients as $\tilde{\alpha}_1 = \alpha_1-\alpha_3$ and $\tilde{\alpha}_2 = \alpha_2+4\alpha_3$. The rewritten bulk action is,
\begin{eqnarray}\label{4Dbulk}
	\hat{S}_{bulk} && =\frac{1}{16\pi G_4} \int d^4x \sqrt{-\hat{g}} \Big( \hat{R} - 2\Lambda -\hat{F}_{AB}\hat{F}^{AB} + \lambda(\tilde{\alpha}_1\hat{R}^2 + \tilde{\alpha}_2\hat{R}_{AB}\hat{R}^{AB} \nonumber\\
	&& + \alpha_3\hat{R}_{GB}^2) \Big).
\end{eqnarray}
The corresponding  boundary action is  %\cite{Cremonini:2009ih},
\begin{eqnarray}\label{4Dbdy}
	\hat{S}_{bdy} &=& \frac{1}{8\pi G_4} \int d^3x \sqrt{-\hat{h}} \Bigg[ \left( 1-\lambda\frac{24}{L^2}\tilde{\alpha}_1 -\lambda\frac{6}{L^2}\tilde{\alpha}_2 \right)\hat{K} + \lambda\tilde{\alpha}_2 (-\hat{K}\hat{F}^2+2\hat{K}\hat{F}^{AB}{\hat{F}}{^C_B}\hat{n}_A \hat{n}_C  \nonumber \\
	&& + 2\hat{K}_{ab}\hat{F}^{aD}{\hat{F}}{^b_D}) + 2\lambda\alpha_3 (\hat{J} - 2\hat{G}^{(3)}_{ab}\hat{K}^{ab}) \Bigg],
\end{eqnarray}
where $\hat{J}$ is the trace of: 
\begin{equation}
	\hat{J}_{ab} = \frac{1}{3} (2\hat{K}\hat{K}_{ac}\hat{K}^c_b + \hat{K}_{cd}\hat{K}^{cd}\hat{K}_{ab} -2\hat{K}_{ac}\hat{K}^{cd}\hat{K}_{db} -\hat{K}^2\hat{K}_{ab}) \label{J_ab}.
\end{equation}
 $\Lambda$ is the four dimensional cosmological constant and given in terms of the four dimensional $AdS$ radius $L$ as 
\begin{equation}
	\Lambda = -\frac{3}{L^2}. 
\end{equation}
For a fixed electrically charged four dimensional solution, we need to impose a boundary condition for the gauge field as follows
\begin{eqnarray}
   \hat{S}_{Maxb} = \frac{1}{4\pi G}\int d^3x \sqrt{-\hat{h}}\hat{n}_C \hat{F}^{CD}\hat{A}_D
\end{eqnarray}
Thus the full action is 
\begin{equation}
    \hat{S} = \hat{S}_{bulk}+\hat{S}_{bdy}+\hat{S}_{Maxb}
\end{equation}
For a near extremal black hole in four dimension, the near horizon geometry takes the form $AdS_2 \times S^2$. We are interested for a higher derivative corrected, electrically charged black hole with charge $Q$. For this solution, the near horizon values of the radii of $S^2$ and $AdS_2$ are given by
\begin{eqnarray}
   \tilde{\Phi}_0 &=&\Phi_0+\lambda(2\tilde{\alpha}_1+\tilde{\alpha}_2)\frac{L^2+3\Phi_0^2}{L^2+6\Phi_0^2}\frac{6\Phi_0}{L^2}\\
   \tilde{L}_2&=&L_2+\lambda (2\tilde{\alpha}_1+\tilde{\alpha}_2)\frac{L^2+3\Phi_0^2}{(L^2+6\Phi_0^2)^{5/2}}6\Phi_0 L
\end{eqnarray}
with $L_2=\sqrt{\frac{\Phi_0^2}{1+\frac{6\Phi_0^2}{L^2}}}$, the uncorrected $AdS_2$ radius. $\Phi_0$ is the uncorrected $S^2$ radius such that
\begin{eqnarray}
   \Phi_0^2 = \frac{L^2}{6}\left(\sqrt{1+\frac{12Q^2}{L^2}}-1\right).
\end{eqnarray}

We find an effective two dimensional action following the  dimensional reduction over a spherically symmetric metric
\begin{eqnarray}
   ds^2 = \tilde{g}_{\alpha\beta}dx^\alpha dx^\beta+\Phi^2(x)(d\theta^2+\sin^2\theta d\phi^2), \quad x^\alpha = t,r
\end{eqnarray}
followed by integrating out the gauge field. One can write a JT-like action by expanding the above mentioned reduced theory around the constant solution $\tilde{\Phi}_0$. At the first order of the expansion, we find the JT-like action, given by \cite{Banerjee:2021vjy}
\begin{equation}\label{dybl}
	S_{bulk} = -\frac{1}{16\pi G_2}\int d^2x \sqrt{g}\big[\phi\left(R+\frac{2}{\tilde{L}_2^2}\right)+\lambda\frac{12\tilde{\alpha}_1\Phi_0^2-\tilde{\alpha}_2L^2}{2\left(L^2+6\Phi_0^2\right)}\left(R^2-\frac{4}{L_2^4}\right)\phi\big]
\end{equation} 
where $\phi$ is the fluctuation given as $\Phi = \tilde{\Phi}_0(1+\phi)$. $G_2$ is the two dimensional Newton's constant related to the $G_4$ as 
$G_2 = \frac{G_4}{8\pi\tilde{\Phi}_0^2}$.

Instead of four dimensional black hole, we could dimensionally reduce our starting five dimensional action over a spherically symmetric metric. In that case we would get a similar JT -like action, since in two dimensions the Riemann curvature has only one independent component. The only significant difference would happen in the potential for the dilation $\phi$. As a higher derivative corrected JT like action reduced from four dimensional theory is already known, we computed the holographic complexity for that model to understand the implications of higher derivative interaction terms in two dimensions.   

The boundary action needed for the well defined variational principle is given by
\begin{eqnarray}
	S_{bdry} = -\frac{1}{8\pi G_2}\int dx \sqrt{h}\left(1+\lambda\frac{2\tilde{\alpha}_2}{\Phi_0^2}-\lambda\frac{24\tilde{\alpha}_1}{L^2}\right)\phi K
\end{eqnarray}

\iffalse
Whereas the boundary counter term is
\begin{eqnarray}
	S_{counter} &=& \frac{1}{8\pi G_2}\int dx \sqrt{h}\left(1+\lambda\frac{2\tilde{\alpha}_2}{\Phi_0^2}-\lambda\frac{24\tilde{\alpha}_1}{L^2}\right) \frac{\phi(r)^2}{\tilde{L}_2\sqrt{f(r)}}\nonumber\\&&
	\times \left(1+\lambda\frac{12\tilde{\alpha}_1\Phi_0^2-\tilde{\alpha}_2L^2}{4\left(L^2+6\Phi_0^2\right)}\left(R^2-\frac{4}{L_2^4}\right)\right)L_2^2
\end{eqnarray}
	\fi

At the zeroth order of the expansion, i.e. evaluating the effective two dimensional action over the constant dilaton solution, we find 
\begin{eqnarray}
    S^0_{bulk} &=&-\frac{1}{32\pi\tilde{\Phi}_0^2 G_2}\int d^2x \sqrt{g}\bigg[\left(\tilde{\Phi}_0^2+\lambda 4\tilde{\alpha}_1+\lambda 4 \alpha_3\right)R\nonumber\\&&
    +\lambda(2\tilde{\alpha}_1+\tilde{\alpha}_2)\bigg\{\frac{24}{L^2}\left(1+\frac{3\Phi_0^2}{L^2}\right)+\frac{2}{\Phi_0^2}+\frac{1}{2}R^2\Phi_0^2\bigg\}\bigg]
    \end{eqnarray}
    \begin{eqnarray}
   S^0_{bdry} = -\frac{1}{16\pi\tilde{\Phi}_0^2 G_2}\int dx\sqrt{h}\left(\tilde{\Phi}_0^2-\lambda 2 \tilde{\alpha}_2+\lambda 4\alpha_3-\lambda(2\tilde{\alpha}_1+\tilde{\alpha}_2)\frac{12\Phi_0^2}{L^2}\right)K
\end{eqnarray}
$S^0_{bulk}+S^0_{bdry}$ is topological and is equal to the  extremal entropy.
Our aim is to compute the holographic complexity of this two dimensional higher derivative corrected theory similarly like previous sections by using CA conjecture.

\subsection{Computation of Holographic Complexity}
    \begin{figure}[!htb]
    \begin{center}
\includegraphics[scale=0.3]{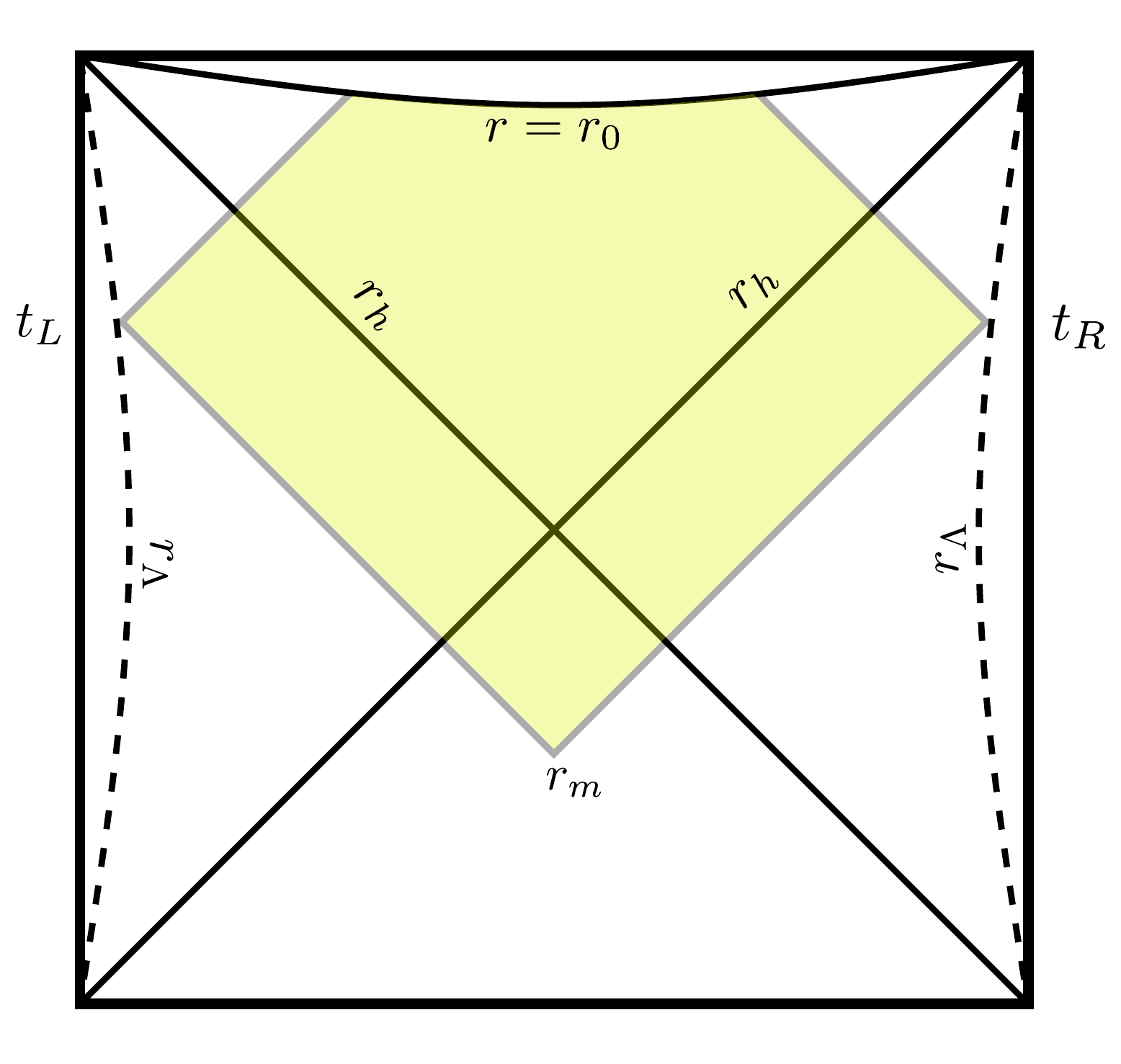}
\end{center}
\label{fig:4}
\caption{Penrose diagram for the JT black hole}
\end{figure}

Equations of motion derived from the higher derivative corrected two dimensional dilaton gravity action \eqref{dybl} are solved by the following $AdS_2$ solutions
\begin{eqnarray}
   ds^2=-f(r)dt^2+\frac{dr^2}{f(r)} , \quad \phi(r)=\frac{r}{\tilde{L}_2}
\end{eqnarray}
where $f(r)=\frac{(r^2-r_h^2)}{\tilde{L}_2^2}$.

The dynamical bulk term $S_{bulk}$ drops to zero on-shell. Thus the only non-trivial bulk contribution to the complexity arise from the topological term $S^0_{bulk}$. In this case, there are two types of boundary of the WDW patch, null-like and space-like \cite{Alishahiha:2018swh}. By choosing an affine parametrization of  the null directions, contribution from the null boundary can be made zero. In this way, the only non-trivial boundary contribution comes from the space-like boundary ($r=0$) where we have introduced a cut-off surface at $r=r_0$ in Fig. 2. After computation of the late time complexity growth rate, we can take $r_0\rightarrow 0$ limit. Thus at late time, contribution from the boundary term $S_{bdry}$ drops to zero since it is polynomial in $r_0$. As mentioned before, on-shell $S^0_{bulk}+S^0_{bdry}$ correspond to the extremal entropy ${\cal S}_0$. Thus it can be shown that the contribution to the late time growth rate of complexity comes from $S^0_{bulk}+S^0_{bdty}$ as ${\cal S}_0 T$ where $T$ is the temperature of the black hole and ${\cal S}_0$ is the extremal entropy. Another piece of non-trivial contribution to $\frac{dC_A}{dt}|_{t\rightarrow\infty}$ comes from the joint term at $r=r_m$, due to both topological and dynamical actions \cite{Alishahiha:2018swh}. Following \cite{Jiang:2018sqj}, we see that the total late time joint term contribution is 
 \begin{eqnarray}
    \frac{dI_{joint}}{dt} = T {\cal S}_n+T {\cal S}_0
 \end{eqnarray}
 where ${\cal S}_n$ is the near extremal part of the entropy. The expression for ${\cal S}_0$ and ${\cal S}_n$ in terms of the four dimensional quantities are
\begin{eqnarray}
   {\cal S}_0 = \frac{\pi \Phi_0^2}{G_4}-\lambda(\tilde{\alpha}_2-2\alpha_3)\frac{2\pi }{G_4}-\lambda(2\tilde{\alpha}_1+\tilde{\alpha}_2)\frac{36\pi\Phi_0^4}{G_4 L^2(L^2+6\Phi_0^2)}
\end{eqnarray}
\begin{eqnarray}
   {\cal S}_n = 4\pi^2\frac{\tilde{L}_2^2\tilde{\Phi}_0}{G_4} \left(1+\lambda\left(\frac{2\tilde{\alpha}_2}{\Phi_0^2}-\frac{24\tilde{\alpha}_1}{L^2}\right)\right)T
\end{eqnarray}
where $G_4$ is the four dimensional Newton constant and $L$ is the $AdS$ length of the four dimensional black hole solution. Here we are ignoring the quantum correction part of the near extremal entropy.

Therefore at late times, the growth rate of the total holographic complexity is
\begin{eqnarray}
  \pi \frac{dC_A}{dt}\bigg|_{t\rightarrow\infty} &=& 2{\cal S}_0 T + {\cal S}_n T\label{jtc}
\end{eqnarray}
 
We can rewrite \eqref{jtc} as,
\begin{eqnarray}
  \pi \frac{dC_A}{dt} &=& 2{\cal S}_{tot} T + \mathcal{O}(T^2)
\end{eqnarray}
where ${\cal S}_{tot}={\cal S}_0+{\cal S}_n$ is the total entropy of the near extremal black hole. Comparing with the earlier works on the complexity of JT gravity \cite{Alishahiha:2018swh,Brown:2018bms}, we see that the late time growth rate of higher derivative corrected JT gravity has the same behaviour as the JT gravity.
%%%
\section{Conclusion}\label{sec4}
In this paper, first, we have studied the holographic complexity of a CFT that is dual to a charged gravitational theory with generic four derivative corrections in five dimensions. We started with the most generic four derivative terms involving metric and the gauge field. We have considered the higher derivative corrections in such a way that they modify not only the metric but also the chemical potential. We compute the holographic complexity for the same action via ``\textit{Complexity = Action}" conjecture. For Einstein's gravity, the late time growth of holographic complexity has an upper bound as  $ \tfrac{d\mathcal{C}_A}{dt}\big|_{t \to \infty} \leq 2M $ and we found that the bound does not hold in generic four derivative gravity. We find that among the interaction terms, specifically the terms $R_{\mu\nu\alpha\beta}R^{\mu\nu\alpha\beta}$ and $R_{\mu\nu\alpha\beta}F^{\mu\nu}F^{\alpha\beta}$ affect the late time growth of the holographic action complexity in such a way that it violates the well known Llyod's bound. However, for many known higher derivative gravity theories, this bound is satisfied unlike this fourth order derivative theory.

It is interesting to note that, the same terms i.e. $R_{\mu\nu\alpha\beta}R^{\mu\nu\alpha\beta}$ and $R_{\mu\nu\alpha\beta}F^{\mu\nu}F^{\alpha\beta}$ violate the KSS bound \cite{Myers:2009ij} on the $\eta/s$ ratio of the dual fluid. This comparison can be verified easily for a specific case when $a_2= -\frac{1}{2} a_1$ and $a_1$. For these choices of the $a_1$ and $a_2$, both the late time Lloyd's bound and KSS bound are violated. It would be interesting to compute the holographic complexity of this higher derivative corrected gravity theory following the CV conjecture and compare with the present result. We have also discussed the late time complexity growth rate of the JT gravity with four derivative correction. We find that in this case the late time growth rate has the same form as it has for two derivative JT gravity. This is due to the fact that the model of higher derivative corrected JT gravity we considered has the same properties like an uncorrected JT gravity. Like the JT gravity, this higher derivative corrected JT gravity can be written as a boundary Schwarzian theory and duly describes the near horizon dynamics of a four dimensional, near extremal black hole in presence of higher derivative corrections.

One other possible direction is to explore the complexity from the dual field theory side. First the idea of computational complexity was proposed by Nielsen et al. \cite{arxiv.quant-ph/0502070, Nielsen_2006}. It can be measured by estimating distances in the manifold of allowed unitary operations required to evolve from a reference state to a target state. The computational complexity can be derived by minimizing distances thus finding the length of the shortest geodesic. This approach was then generalized for free quantum field theories in \cite{Jefferson:2017sdb}. However to understand the holographic complexity of AdS black hole spacetime in terms of boundary theory quantities one have to understand the notion of complexity in CFTs. There has been a recent proposal for finding complexity in 2D CFTs \cite{Caputa:2018kdj}, wherein the subset of unitary symmetry gates are constructed from the energy momentum tensor of 2D CFTs. It will be very interesting to find out whether similar unitary gates can be constructed from the fluid stress energy tensor and the KSS bound can be directly related with the Llyod bound of computational complexity.

\section*{ Acknowledgments}We would like to thank Stefano Baiguera, Nabamita Banerjee, Suvankar Dutta, Debangshu Mukherjee and Muktajyoti Saha for helpful discussions and comments on our initial draft. The work of TM is supported by a Simons Foundation Grant Award ID 509116 and by the South African Research Chairs initiative of the Department of Science and Technology and the National Research Foundation. This work of AM was supported by IISER Bhopal and the Ministry of Science and Technology, National Center for Theoretical Sciences of Taiwan.
\bibliography{ssbib}

\end{document}